\documentstyle[]{article}
\def\bea{\begin{eqnarray}}
\def\eea{\end{eqnarray}}
\def\nn{\nonumber}

\def\beq{\begin{equation}}
\def\eeq{\end{equation}}
\def\ba{\beq\new\begin{array}{c}}
\def\ea{\end{array}\eeq}
\def\be{\ba}
\def\ee{\ea}

\def\Tr{{\rm Tr}}

\def\2{{1\over 2}}
\def\f{1\over}
\makeatletter
\newdimen\normalarrayskip              
\newdimen\minarrayskip                 
\normalarrayskip\baselineskip
\minarrayskip\jot
\newif\ifold             \oldtrue            \def\new{\oldfalse}
\def\arraymode{\ifold\relax\else\displaystyle\fi} 
\def\eqnumphantom{\phantom{(\theequation)}}     
\def\@arrayskip{\ifold\baselineskip\z@\lineskip\z@
     \else
     \baselineskip\minarrayskip\lineskip2\minarrayskip\fi}
\def\@arrayclassz{\ifcase \@lastchclass \@acolampacol \or
\@ampacol \or \or \or \@addamp \or
   \@acolampacol \or \@firstampfalse \@acol \fi
\edef\@preamble{\@preamble
  \ifcase \@chnum
     \hfil$\relax\arraymode\@sharp$\hfil
     \or $\relax\arraymode\@sharp$\hfil
     \or \hfil$\relax\arraymode\@sharp$\fi}}
\def\@array[#1]#2{\setbox\@arstrutbox=\hbox{\vrule
     height\arraystretch \ht\strutbox
     depth\arraystretch \dp\strutbox
     width\z@}\@mkpream{#2}\edef\@preamble{\halign
\noexpand\@halignto
\bgroup \tabskip\z@ \@arstrut \@preamble \tabskip\z@ \cr}%
\let\@startpbox\@@startpbox \let\@endpbox\@@endpbox
  \if #1t\vtop \else \if#1b\vbox \else \vcenter \fi\fi
  \bgroup \let\par\relax
  \let\@sharp##\let\protect\relax
  \@arrayskip\@preamble}
\def\eqnarray{\stepcounter{equation}%
              \let\@currentlabel=\theequation
              \global\@eqnswtrue
              \global\@eqcnt\z@
              \tabskip\@centering
              \let\\=\@eqncr
              $$%
 \halign to \displaywidth\bgroup
    \eqnumphantom\@eqnsel\hskip\@centering
    $\displaystyle \tabskip\z@ {##}$%
    \global\@eqcnt\@ne \hskip 2\arraycolsep
         $\displaystyle\arraymode{##}$\hfil
    \global\@eqcnt\tw@ \hskip 2\arraycolsep
         $\displaystyle\tabskip\z@{##}$\hfil
         \tabskip\@centering
    &{##}\tabskip\z@\cr}
\begingroup\ifx\undefined\newsymbol \else\def\input#1 {\endgroup}\fi
\input amssym.def \relax
\input amssym
\newfont{\hr}{msbm10}
\newfont{\ams}{msam10}
\font\teneufm=cmmib10
\font\seveneufm=cmmib7
\font\fiveeufm=cmmib5
\def\bfit#1{{\textfont1=\teneufm\scriptfont1=\seveneufm
\scriptscriptfont1=\fiveeufm
\mathchoice{\hbox{$\displaystyle#1$}}{\hbox{$\textstyle#1$}}
{\hbox{$\scriptstyle#1$}}{\hbox{$\scriptscriptstyle#1$}}}}

\def\balpha{{\bfit\alpha}}

\def\bsigma{{\bfit\sigma}}

\begin{document}

\begin{titlepage}
\setcounter{footnote}0
\begin{center}
\hfill ITEP/TH-9/96\\
\hfill FIAN/TD-8/96\\
\vspace{0.3in}
\bigskip
\bigskip
{\LARGE\bf A Note on Spectral Curve for the Periodic Homogeneous $XYZ$-Spin
Chain
\footnote{Contribution to the volume "Problems in Modern Theoretical
Physics" in honour of the 60th birthday of Professor A.T.Filippov}
}
\\
\bigskip
\bigskip
\bigskip
{\Large A.Gorsky
\footnote{E-mail address: gorsky@vxitep.itep.ru, sasha@rhea.teorfys.uu.se}
$^{\dag}$,
A.Marshakov
\footnote{E-mail address:
mars@lpi.ac.ru, andrei@rhea.teorfys.uu.se, marshakov@nbivax.nbi.dk}$^{\ddag}$,
A.Mironov
\footnote{E-mail address:
mironov@lpi.ac.ru, mironov@grotte.teorfys.uu.se}$^{\ddag}$,
A.Morozov
\footnote{E-mail address:
morozov@vxdesy.desy.de}
$^{\dag}$}
\\
\bigskip
$\phantom{gh}^{\dag}${\it ITEP, Moscow ~117 259, Russia}\\
$\phantom{gh}^{\ddag}${\it Theory Department, P. N. Lebedev Physics
Institute, Leninsky prospect 53, Moscow, ~117924, Russia\\
and ITEP, Moscow ~117259, Russia}
\end{center}
\bigskip \bigskip

\begin{abstract}
We discuss the construction of the spectral curve and the
action integrals for the ``elliptic" $XYZ$ spin chain of
the length $N_c$. Our analysis can reflect the integrable structure
behind the ``elliptic" ${\cal N}=2$ supersymmetric QCD with $N_f=2N_c$ .

\end{abstract}

\end{titlepage}

\newpage
\setcounter{footnote}0

In recent paper \cite{GMMM} we argued that the integrable
counterpart \cite{GKMMM}-\cite{NT2}
of the low-energy $4d$ ${\cal N}=2$ SUSY Yang-Mills
theory with $N_f$ fundamental matter hypermultiplets
\cite{SW1}-\cite{FuMa} can be searched among integrable spin chains.
A reasonable suggestion was made in \cite{GMMM} about the
``rational'' case of $N_f<2N_c$ (see also \cite{M,AN}). However,
the story can not be complete without studying the most
intriguing ``elliptic'' case of $N_f=2N_c$, when the $4d$ theory
is UV-finite and possesses an extra {\it dimensionless}
parameter: the UV non-abelian coupling constant
$\tau = \frac{8\pi i}{e^2} + \frac{\theta}{\pi}$.
Before addressing to the physical problems, some
preliminary technical work should be done, and we start such work in
the present paper by analyzing relevant aspects of the
theory of $XYZ$ spin chains. The $4d$ interpretation of this theory
will not be directly addressed here and postponed to a separate paper.

The main ingredients of interest for us are the spectral curve
${\cal C}$, associated with the spin chain model, and the ``action
integrals" $a_i = \oint_{A_i} dS$, $a_D^i = \oint_{B^i}dS$, or
the moduli-dependence of the cohomology class of the ``generating"
1-form $dS$ defined by the property
\be\label{basic}
\frac{\partial dS}{\partial \{ moduli\}}\ \cong\
\hbox{holomorphic}\ \hbox{1-form}
\ee
The main information on the $XYZ$ models can be found in refs.
\cite{Skl}-\cite{KriZa} (see also \cite{JapBob} for its continuum limit).

\section{Toda chain: $N_c\times N_c$ versus $2\times 2$
representation}

We begin our analysis from the simplest Toda-chain model, which in the
framework of the Seiberg-Witten solutions
corresponds to the $4d$ {\it pure} gauge ${\cal N}=2$ supersymmetric Yang-Mills
theory. The periodic problem in this model can be formulated in two different
ways, which will be further deformed into two different directions. These
deformations
are hypothetically related to the two different deformations
of the $4d$ theory by adding the adjoint and fundamental matter
${\cal N}=2$ hypermultiplets correspondingly.

The Toda chain system can be defined by the equations of motion
\be
\frac{\partial q_i}{\partial t} = p_i \ \ \ \ \
\frac{\partial p_i}{\partial t} = e^{q_{i+1} -q_i}- e^{q_i-q_{i-1}}
\ee
where one assumes (for the periodic problem with the ``period" $N_c$) that
$q_{i+N_c} = q_i$ and
$p_{i+N_c} = p_i$. It is an integrable system, with $N_c$
Poisson-commuting Hamiltonians, $h_1^{TC} = \sum p_i$, $h_2^{TC} =
\sum\left(\frac{1}{2}p_i^2 + e^{q_i-q_{i-1}}\right)$, $\ldots$.  The
generation function for these Hamiltonians can be written in terms of a Lax
operator and as was already mentioned, the Toda chain possesses two
essentially different formulations of this kind.

In the first version (which can be considered as a limiting case of
Hitchin system \cite{Hit}), the Lax operator is the $N_c\times N_c$
matrix-valued 1-form,
\be\label{LaxTC}
{\cal L}^{TC}(w)\frac{dw}{w} =
\left(\begin{array}{ccccc}
 p_1 & e^{{1\over 2}(q_1-q_2)} & 0 & & we^{{1\over 2}(q_1-q_{N_c})}\\
e^{{1\over 2}(q_2-q_1)} & p_2 & e^{{1\over 2}(q_2 - q_3)} & \ldots & 0\\
0 & e^{{1\over 2}(q_3-q_2)} & p_3 & & 0 \\
 & & \ldots & & \\
\frac{1}{w}e^{{1\over 2}(q_{N_c}-q_1)} & 0 & 0 & & p_{N_c}
\end{array} \right)\frac{dw}{w}
\ee
defined on the two-punctured sphere. The Poisson brackets
$\{p_i,q_j\} = \delta_{ij}$ imply that
\be
\left\{ {\cal L}^{TC}(w)\stackrel{\otimes}{,}
{\cal L}^{TC}(w')\right\} =
\left[ {\cal R}(w,w'), {\cal L}^{TC}(w)\otimes {\bf 1} + {\bf 1}\otimes
{\cal L}^{TC}(w')\right]
\ee
with the {\it numeric} trigonometric
${\cal R}$-matrix \cite{Jimbo},
and the eigenvalues of the Lax operator defined from the spectral
equation
\be\label{SpeC}
\det_{N_c\times N_c}\left({\cal L}^{TC}(w) -
\lambda\right) = 0
\ee
are Poisson-commuting with each other. Moreover, the set of
the action integrals along the cycles $\hat\gamma_a $ in the $2N_c$-dimensional
phase space of the system can be reproduced by the integrals of the
generating ``eigenvalue" 1-form $dS^{TC} \cong \lambda\frac{dw}{w}$ along
some non-contractable cycles $\gamma_a $
on the spectral curve ${\cal C}^{TC}$
defined by eq.(\ref{SpeC}):
\be
I_{\gamma_a} =\oint_{\hat\gamma_a}
\sum_{i=1}^{N_c} p_idq_i = \oint_{\gamma_a} dS
\ee
The generating differential
$dS$ plays an essential role in the theory of the (finite - gap) integrable
systems.
It gives rise to the symplectic form on the phase space of the finite-gap
solutions, describes their Whitham-like deformations \cite{KriW,GKMMM,M}
and
allows one to find explicitely the action-angle variables in the framework of
so-called ``separation of variables" \cite{Skl} and Hitchin formalism
\cite{Hit} (see also \cite{NekRub}).
Substituting the explicit expression (\ref{LaxTC}) into (\ref{SpeC}),
one gets \cite{KriDu}:
\be\label{fsc-Toda}
w + \frac{1}{w} = 2P_{N_c}(\lambda )
\ee
where $P_{N_c}(\lambda )$ is a polynomial of degree $N_c$,
with the coefficients being the Schur polynomials of the Hamiltonians
$h_k = \sum_{i=1}^{N_c} p_i^k + \ldots$:
\be
P_{N_c}(\lambda ) = \sum_{k=0}^{N_c}
{\cal S}_{N_c-k}(h) \lambda ^{N_c} = \nn \\
= \left( \lambda ^{N_c} + h_1 \lambda ^{N_c-1} +
\frac{1}{2}(h_2-h_1^2)\lambda ^{N_c-2} + \ldots \right)
\ee
The spectral equation depends only on the mutually Poisson-commuting
combinations of the dynamical variables --
the Hamiltonians --
parametrizing (a subspace in the) moduli space of the complex structures of
the hyperelliptic curves ${\cal C}^{TC}$ of genus $N_c - 1$.

An alternative description of the same system involves (a chain of) $2\times
2$ matrices \cite{FT},
\be\label{LTC}
L^{TC}_i(\lambda) =
\left(\begin{array}{cc} p_i + \lambda & e^{q_i} \\ e^{-q_i} & 0
\end{array}\right), \ \ \ \ \ i = 1,\dots ,N_c
\ee
obeying the {\it
quadratic} $r$-matrix Poisson relations \cite{Skl} \be\label{quadrP} \left\{
L^{TC}_i(\lambda)\stackrel{\otimes}{,}L_j^{TC}(\lambda')\right\} =
\delta_{ij} \left[ r(\lambda - \lambda'), L^{TC}_i(\lambda)\otimes
L^{TC}_j(\lambda')\right]
\ee
with the ($i$-independent!) numerical rational
$r$-matrix satisfying the classical Yang-Baxter
equation $r(\lambda) =
\frac{1}{\lambda } \sum_{a=1}^3 \sigma_a\otimes \sigma^a$.
As a consequence, the transfer matrix (generally defined for the
inhomogeneous lattice with inhomogenities $\lambda_i$'s)
\be\label{monomat}
T_{N_c}(\lambda) =
\prod_{1\ge i\ge N_c}^{\curvearrowleft} L_i(\lambda - \lambda_i)
\ee
satisfies the same Poisson relation
\be
\left\{ T_{N_c}(\lambda)\stackrel{\otimes}{,}T_{N_c}(\lambda')\right\}
= \left[ r(\lambda - \lambda'), T_{N_c}(\lambda)\otimes
T_{N_c}(\lambda')\right]
\ee
and the integrals of motion of the Toda chain are generated
by another form of spectral equation
\be\label{spec}
\det_{2\times 2}\left( T^{TC}_{N_c}(\lambda )
- w\right) = w^2 - w\Tr T^{TC}_{N_c}(\lambda ) + 1 = 0
\ee
or
\be\label{specTC}
w + \frac{1}{w} = \Tr T^{TC}_{N_c}(\lambda)
\ee
(We used the fact that
$\det_{2\times 2} L^{TC}(\lambda) = 1$ leads to $\det_{2\times 2} T_{N_c}^{TC}
(\lambda) = 1$.) The r.h.s. of (\ref{specTC}) is a polynomial of degree $N_c$
in $\lambda$, with the coefficients being the integrals of motion since
\be\label{trcom}
\left\{ \Tr T_{N_c}(\lambda ), \Tr T_{N_c}(\lambda' )\right\} =
\Tr \left\{ T_{N_c}(\lambda )\stackrel{\otimes}{,}T_{N_c}(\lambda' )\right\} =
\nn \\
= \Tr \left[ r(\lambda - \lambda' ), T_{N_c}(\lambda )\otimes
T_{N_c}(\lambda' )\right] = 0
\ee
For the particular choice of $L$-matrix (\ref{LTC}), the inhomogenities
of the chain, $\lambda_i$, can be absorbed into the redefinition
of the momenta $p_i \rightarrow p_i - \lambda_i$.
It is possible to establish a straightforward relation between two
representations (\ref{SpeC}) and (\ref{spec}) (see \cite{GMMM} for
details).

In what follows we consider possible elliptic deformations of
two Lax representations of the Toda chain. The deformation of the
$N_c\times N_c$ representation provides the Calogero-Moser model
(discussed in the context of the Seiberg-Witten approach in \cite{intcal}),
while the deformation of the spin-chain $2\times 2$ representation gives rise
to the Sklyanin $XYZ$ model.

\section{Elliptic deformation of the $N_c\times N_c$ representation:
the Calogero-Moser model \label{Cal}}

The $N_c\times N_c$ matrix-valued
Lax 1-form for the $GL(N_c)$ Calogero system is \cite{KriCal}
\be\label{LaxCal}
{\cal L}^{Cal}(\xi)d\xi =
\left({\bf pH} + \sum_{\balpha}F({\bf {\goth q}\balpha}|\xi)
E_{\balpha}\right)d\xi = \nn \\
= \left(\begin{array}{cccc}
 p_1 & F({\goth q}_1-{\goth q}_2|\xi) & \ldots &
F({\goth q}_1 - {\goth q}_{N_c}|\xi)\\
F({\goth q}_2-{\goth q}_1|\xi) & p_2 & \ldots &
F({\goth q}_2-{\goth q}_{N_c}|\xi)\\
 & & \ldots  & \\
F({\goth q}_{N_c}-{\goth q}_1|\xi) &
F({\goth q}_{N_c}-{\goth q}_2|\xi)& \ldots &p_{N_c}
\end{array} \right)d\xi
\ee
The function $F({\goth q}|\xi) = \frac{g}{\omega}
\frac{\sigma({\goth q}+\xi)}{\sigma({\goth q})\sigma(\xi)}
e^{\zeta({\goth q})\xi}$ and, thus, the Lax operator ${\cal L}(\xi)d\xi$ is
defined on the elliptic curve $E(\tau)$ (complex torus
with periods $\omega, \omega '$ and modulus $\tau = \frac{\omega '}{\omega}$).
The Calogero coupling constant is $\frac{g^2}{\omega^2} \sim m^2$,
where in the $4d$
interpretation $m$ plays the role of the mass of the adjoint matter
${\cal N}=2$ hypermultiplet breaking ${\cal N}=4$ SUSY down to ${\cal N}=2$.

The spectral curve ${\cal C}^{Cal}$ for the $GL(N_c)$ Calogero system is
given by:
\be\label{fscCal}
\det_{N_c\times N_c} \left({\cal L}^{Cal}(\xi) - \lambda\right) = 0
\ee
The periods $a_i$ and $a_i^D$ are again the integrals of the
generating 1-differential
\be
dS^{Cal} \cong \lambda d\xi
\ee
along the non-contractable contours on ${\cal C}^{Cal}$.
Integrability of the Calogero system is implied by the Poisson
structure of the form
\be\label{lin-r}
\left\{ {\cal L}(\xi)\stackrel{\otimes}{,}{\cal L}(\xi ')\right\}
= \left[ {\cal R}_{12}^{Cal}(\xi ,\xi '),\ {\cal L}(\xi)\otimes {\bf 1}\right]
-
\left[ {\cal R}_{21}^{Cal}(\xi ,\xi '),\ {\bf 1} \otimes {\cal L}(\xi') \right]
\ee
with the {\it dynamical} elliptic ${\cal R}$-matrix \cite{RCal},
guaranteeing that the eigenvalues of the matrix ${\cal L}$ are in involution.

In order to recover the Toda-chain system, one takes the double-scaling
limit \cite{Ino}, when $g \sim m$ and $-i\tau$ both go
to infinity (and
${\goth q}_i-{\goth q}_j={\2}\left[(i-j)\log g +(q_i-q_j)\right]$)
so that the dimensionless coupling $\tau$ gets
substituted by a dimensional parameter $\Lambda^{N_c} \sim
m^{N_c}e^{i\pi\tau}$. In this limit,
the elliptic curve $E(\tau)$ degenerates into the (two-punctured) Riemann
sphere with coordinate $w = e^{\xi }e^{i\pi\tau}$ so that
\be
dS^{Cal} \rightarrow dS^{TC} \cong \lambda\frac{dw}{w}
\ee
The Lax operator of the Calogero system turns into that of the
$N_c$-periodic Toda chain (\ref{LaxTC}):
\be
{\cal L}^{Cal}(\xi )d\xi \rightarrow {\cal L}^{TC}(w)\frac{dw}{w}
\ee
and the spectral curve acquires the form of (\ref{SpeC}).
In contrast to the Toda case, (\ref{fscCal}) can {\it not} be rewritten in
the form (\ref{fsc-Toda}) and peculiar $w$-dependence of the spectral
equation (\ref{SpeC}) is not preserved by embedding of Toda into Calogero-Moser
particle
system. However, the form (\ref{fsc-Toda}) can be naturally preserved by
the alternative deformation of the Toda-chain system when it
is considered as (a particular case of) a spin-chain model.

To deal with the ``elliptic" deformations of the Toda chain below, we will use
a non-standard normalization of the Weierstrass
$\wp $-function defined by
\be\label{wpmod}
\wp(\xi |\tau) = \sum_{m,n = -\infty}^{+\infty}
\frac{1}{(\xi + m + n\tau)^2} - {\sum_{m,n=-\infty}^{+\infty}}'
\frac{1}{(m+n\tau)^2}
\ee
so that it is double periodic in $\xi $ with periods $1$ and
$\tau = {\omega '\over \omega}$ (that differs from a standard definition
by a factor of $\omega^{-2}$ and by
rescaling $\xi \rightarrow \omega\xi$). According to (\ref{wpmod}), the
values of $\wp(\xi |\tau)$ in the half-periods, $e_a = e_a(\tau )$, $a=1,2,3$,
are also the functions only of $\tau$ -- again differing by a factor
of $\omega^{-2}$ from the conventional definition.

The complex torus $E(\tau )$ can be defined as ${\bf C}/{\bf Z}\oplus
\tau{\bf Z}$
with a
``flat" co-ordinate $\xi$ defined $modulo(1,\tau)$.
Alternatively, any torus (with a marked point) can be described as
elliptic curve
\be\label{ell}
y^2 = (x -  e_1)(x - e_2)(x - e_3) \nn \\ x
= \wp(\xi ) \ \ \  \ \ y = \frac{1}{2}\wp'(\xi )
\ee
and the canonical
holomorphic 1-differential is
\be
d\xi = 2\frac{dx}{y}
\ee
There are three
interesting degeneration limits:\\
\noindent
-- \underline{rational limit}:
both periods $\omega, \omega ' \rightarrow \infty$, $\xi$ scales as $\xi =
\omega^{-1}\zeta $ with $\tau = {\omega ' \over \omega}$ and $\zeta $ remain
finite. Then:
\be\label{rat}
x = \wp(\xi) = \frac{\omega^2}{\zeta ^2}(1 +
o(\omega^{-1}))  \ \ \ \ y = \frac{1}{2}\wp'(\xi ) = -\frac{\omega^3}{\zeta
^3}(1 + o(\omega^{-1}))
\ee
In two other limits $\tau \rightarrow +i\infty$,
i.e.  $q = e^{i\pi\tau} \rightarrow 0$.  \\
\noindent
--
\underline{trigonometric limit}: $\xi$ remains finite as $q\rightarrow 0$
\be\label{tri}
x = \wp(\xi) = -{1\over 3} + \frac{1}{\sin^2 \pi\xi} + o(q)  \ \ \
y = \frac{1}{2} \wp'(\xi) = -\pi\frac{\cos \pi\xi}{\sin^3\pi\xi} + o(q)
\ee
\noindent
-- \underline{double-scaling limit}: $\xi = \log (qw)$, the branch points
\be\label{brpo}
e_{1,2} \rightarrow -\frac{1}{3} \pm 8q + o(q^2) \ \ \ \ \
e_3 \rightarrow  +\frac{2}{3} + o(q^2)
\ee
and
\be\label{ds}
x = \wp(\xi ) = -\frac{1}{3} + 4q(w + w^{-1}) + o(q^2)  \ \ \ \
y = \frac{1}{2}\wp'(\xi ) = 4q(w - w^{-1}) + o(q^2)
\ee
and
\be
d\xi = \frac{dw}{w}(1 + {\cal O}(q))
\ee
In the simplest example of $N_c=2$,
the spectral curve ${\cal C}^{Cal}$ has genus 2. Indeed,
in this particular case, eq.(\ref{fscCal}) turns into
\be
\lambda ^2 = h_2 - {g^2\over \omega^2}\wp(\xi ) = h_2 - {g^2\over \omega^2}x
\ee
This equation says that with any value of $x$ one
associates two points of ${\cal C}^{Cal}$,
$\lambda = \pm\sqrt{h_2 - \frac{g^2}{\omega^2}x}$, i.e.
it describes ${\cal C}^{cal}$ as a double covering of the
elliptic curve $E(\tau)$ ramified at the points
$x = \left( {\omega\over g}\right)^2h_2$
and $x = \infty$. In fact, since $x$ is an elliptic
coordinate on $E(\tau)$ (when elliptic curve is
treated as a double covering over the Riemann sphere $CP^1$),
$x = \left( {\omega\over g}\right)^2h_2$ corresponds to a {\it pair} of
points on $E(\tau)$ distinguished by the sign of $y$. This would be true
for $x = \infty$
as well, but $x = \infty$ is one of the branch points in
our parametrization (\ref{ell}) of $E(\tau)$. Thus, the {\it two} cuts
between $x = \left( {\omega\over g}\right)^2h_2$ and $x=\infty$ on every
sheet of
$E(\tau)$ touching at the common end at $x=\infty$ become
a {\it single} cut between $\left(\left( {\omega\over g}\right)^2h_2, +\right)$
and $\left(\left( {\omega\over g}\right)^2h_2, -\right)$. Therefore, we can
consider the spectral
curve ${\cal C}^{Cal}$ as two tori $E(\tau)$ glued along one cut, i.e.
${\cal C}^{Cal}_{N_c=2}$ is a curve of genus 2.

Analytically the curve ${\cal C}^{Cal}$ for $N_c = 2$ can be described by the
pair of equations:
\be\label{caln2}
y^2 = \prod_{a=1}^3 (x - e_a),
\nn \\
\lambda ^2 = h_2 - \frac{g^2}{\omega^2} x
\ee
Occasionally, it turns out to be a hyperelliptic curve (for $N_c = 2$ only!)
after substituting in (\ref{caln2}) $x$ from the second equation to the
first one.

Two holomorphic 1-differentials on ${\cal C}^{Cal}$ can be chosen to be
\be\label{holn2}
v = \frac{dx}{y} \sim \frac{\lambda d\lambda}{y}
\ \ \ \ \ \ \
V =  \frac{dx}{y\lambda}\sim \frac{d\lambda }{y}
\ee
so that
\be
dS \cong \lambda d\xi =
\sqrt{h_2 - \frac{g^2}{\omega^2}\wp(\xi )}d\xi =
\frac{dx}{y}\sqrt{h_2 - \frac{g^2}{\omega^2}x}
\ee
It is easy to check the basic property (\ref{basic}):
\be
\frac{\partial dS}{\partial h_2} \cong \frac{1}{2}\frac{dx}{y\lambda}
\ee
The fact that only one of two holomorphic 1-differentials (\ref{holn2})
appears at the r.h.s. is related to their different parity
with respect to the ${\bf Z}_2\otimes {\bf Z}_2$ symmetry of ${\cal C}^{Cal}$:
$y \rightarrow -y$ and $\lambda \rightarrow -\lambda$.
Since $dS$ has certain parity, its
integrals along the two of four elementary non-contractable cycles
on ${\cal C}^{Cal}$ automatically vanish leaving only two
non-vanishing quantities $a$ and $a_D$, as necessary for
the $4d$ interpretation \cite{intcal}. Moreover, these two
non-vanishing
integrals can be actually evaluated in terms of the ``reduced"
genus-{\it one} curve
\be
Y^2 = (y\lambda)^2 = \left(h_2 - \frac{g^2}{\omega^2}x\right)
\prod_{a=1}^3 (x - e_a),
\ee
with $dS \cong \left(h_2 - \frac{g^2}{\omega^2}x\right)\frac{dx}{Y}$.
Since now $x = \infty$ is no longer a ramification point, $dS$
obviously has simple poles at $x = \infty$ (at two points on the
two sheets of ${\cal C}^{Cal}_{reduced}$) with the residues
$\pm\frac{g}{\omega} \sim \pm m$.

The opposite limit of the Calogero-Moser system with vanishing coupling
constant $g^2 \sim m^2 \rightarrow 0$ corresponds to the ${\cal N}=4$ SUSY
Yang-Mills theory with identically vanishing $\beta $-function.
The corresponding integrable
system is a collection of {\it free} particles and the generating differential
$dS \cong \sqrt{h_2}\cdot d\xi $ is just a {\it holomorphic} differential on
$E(\tau )$.

\section{Spin-chain (magnetic) models and Sklyanin algebra. \label{Spin}}

This class of integrable models is based on the quadratic Poisson structure
(\ref{quadrP}) for an $n\times n$ matrix $L(\lambda)$,
which implies the existence of the monodromy matrix (\ref{monomat})
with the Poisson-commuting eigenvalues.

The spectral curve for the {\it periodic inhomogeneous} spin chain is given
by:
\be\label{fsc-SCh}
\det_{n\times n}\left(T_{N_c}(\lambda) - \tilde w\right) = 0
\ee
and the generating 1-differential is
\be
dS \cong \lambda\frac{dw}{w}
\nn \\
w = \tilde w\cdot \det T_{N_c}(\lambda )^{-1/n}
\ee
In the particular case of $n=2$ ($sl(2)$ spin chains), the spectral
equation (\ref{fsc-SCh}) acquires the form:
\be\label{fsc-sc2}
\tilde w + \frac{\det_{2\times 2} T_{N_c}(\lambda)}{\tilde w} =
 \Tr T_{N_c}(\lambda )
\ee
and this peculiar form of $w$-dependence suggests \cite{M,GMMM} that
the periodic $sl(2)$ spin chains are related to the solution to the
${\cal N} = 2$ supersymmetric QCD.

The most general theory of this sort is known as Sklyanin $XYZ$ spin chain
with the elementary $L$-operator defined on the elliptic curve
$E(\tau)$ and is explicitly given by (see \cite{FT} and references therein):
\be\label{39}
L^{Skl}(\xi) = S^0{\bf 1} + i\frac{g}{\omega}\sum_{a=1}^3 W_a(\xi)S^a\sigma_a
\ee
where
\be
W_a(\xi) = \sqrt{e_a - \wp\left({\xi}|\tau\right)} =
i\frac{\theta'_{11}(0)\theta_{a+1}\left({\xi}\right)}{\theta_{a+1}(0)
\theta_{11}\left({\xi}\right)}\\
\theta_2\equiv\theta_{01},\ \ \ \theta_3\equiv\theta_{00},
\ \ \ \theta_{4}\equiv\theta_{10}
\ee
Let us note that our spectral parameter $\xi$ is connected with the standard
one $u$ \cite{FT} by the relation $u=2K\xi$, where
$K\equiv\int_0^{{\pi\over 2}}{dt\over\sqrt{1-k^2\sin^2t}}={\pi\over
2}\theta_{00}^2(0)$, $k^2\equiv{e_1-e_2\over e_1-e_3}$
so that $K\to{\pi\over 2}$ as $q\to 0$. This factor results into additional
multiplier $\pi$ in the trigonometric functions in the limiting cases below.

The Lax operator (\ref{39})
satisfies the Poisson relation (\ref{quadrP}) with the
numerical {\it elliptic} $r$-matrix $r(\xi)={i{g\over\omega}}\sum_{a=1}^3
W_a(\xi)\sigma_a\otimes\sigma_a$, which
implies that $S^0, S^a$ form the (classical)
Sklyanin algebra \cite{Skl1,Skl}:
\be\label{sklyal}
\left\{S^a, S^0\right\} = 2i\left(\frac{g}{\omega}\right)^2
\left(e_b - e_c\right)S^bS^c
\nn \\
\left\{S^a, S^b\right\} = 2iS^0S^c
\ee
with the obvious notation: $abc$ is the triple $123$ or its cyclic
permutations.

The coupling constant  ${g\over \omega}$ can be eliminated by simultaneous
rescaling of the $S$-variables and the symplectic form:
\be
S^a = \frac{\omega}{g}\hat S^a \ \ \ \
S^0 = \hat S^0\ \ \ \
\{\ , \ \} \rightarrow \frac{g}{\omega}\{\ ,\ \}
\ee
Then
\be
L(\xi) = \hat S^0 {\bf 1} + i\sum_{a=1}^3 W_a(\xi)\hat S^a\sigma_a
\ee
\be\label{sklyaln}
\left\{ \hat S^a,\ \hat S^0\right\} = 2i
\left(e_b - e_c\right) \hat S^b\hat S^c \nn \\
\left\{ \hat S^a,\ \hat S^b\right\} = 2i\hat S^0\hat S^c
\ee
In parallel with (\ref{rat})-(\ref{ds}), one can distinguish three interesting
limits of the Sklyanin algebra.\\
--- \underline{rational limit}.
Both $\omega , \omega '\rightarrow\infty $, then
(\ref{sklyal}) turns into
\be\label{ratsklya}
\{ S^a,S^0\} = 0
\nn \\
\{ S^a,S^b\} = 2i\epsilon ^{abc}S^0S^c
\ee
i.e. $S^0$ itself becomes a Casimir operator (constant), while the remaining
$S^a$ form a classical angular-momentum (spin) vector. The corresponding
\be\label{laxrat}
L_{XXX}(\zeta ) = {\bf 1} - {g\over\zeta}{\bf S\cdot\bsigma}
\ee
describes the $XXX$ spin chain considered in detail in \cite{GMMM} with
the rational $r$-matrix.\\
--- \underline{trigonometric limit}.
As $\tau\rightarrow +i\infty$ or $q\rightarrow 0$, the Sklyanin algebra
(\ref{sklyaln}) transforms to
\be\label{trisklya}
\{ \hat S^3,\hat S^0\} = 32iq\hat S^1 \hat S^2 +{\cal O}(q)\rightarrow 0
\nn \\
\{\hat S^1,\hat S^0\} =-2i\hat S^2\hat S^3 + {\cal O}(q)
\nn \\
\{\hat S^2,\hat S^0\} = 2i\hat S^3\hat S^1 + {\cal O}(q)
\nn \\
\{\hat S^1,\hat S^2\} =2i\hat S^0\hat S^3 + {\cal O}(q)
\nn \\
\{\hat S^1,\hat S^3\} =  -2i\hat S^0\hat S^2 + {\cal O}(q)
\nn \\
\{\hat S^2,\hat S^3\} = 2i\hat S^0\hat S^1 + {\cal O}(q)
\ee
The corresponding Lax matrix is
\be\label{laxxxz}
L_{XXZ} = \hat S^0{\bf 1}-{\f \sin\pi\xi}\left(\hat S^1\sigma_1+
\hat S^2\sigma_2+\cos\pi\xi\hat S^3\sigma_3\right)
\ee
and $r$-matrix
\be\label{trigrmatrix}
r(\xi)={i\over\sin\pi\xi}
\left(\sigma_1\otimes\sigma_1+\sigma_2\otimes\sigma_2+
\cos\pi\xi\sigma_3\otimes\sigma_3\right)
\ee
--- \underline{double-scaling limit}. Using (\ref{brpo}) and (\ref{ds}), we
find that
\be\label{predss}
\sqrt{e_{1,2} - \wp \left({\xi}\right)}
=2\sqrt{q}\sqrt{w + {1\over w} \pm 2} +{\cal O}(q)
=2\sqrt{q}\left(\sqrt{w} \pm {1\over\sqrt{w}}\right) + {\cal O}(q)
\nn \\
\sqrt{e_3 - \wp \left({\xi} \right)} =1 +{\cal O}(q)
\ee
and, therefore, the Sklyanin algebra (\ref{sklyaln}), after the rescaling
$\hat S^{1,2}= {\f 4\sqrt{q}}\bar S^{1,2}$, acquires the form
\be\label{sklyads}
\{ \bar S^3,\bar S^0\} = 2i\bar S^1 \bar S^2
\nn \\
\{\bar S^1,\bar S^0\} =-2i\bar S^2\bar S^3 + {\cal O}(q)
\nn \\
\{\bar S^2,\bar S^0\} = 2i\bar S^3\bar S^1 + {\cal O}(q)
\nn \\
\{\bar S^1,\bar S^2\} =32iq\bar S^0\bar S^3 +{\cal O}(q)\rightarrow 0
\nn \\
\{\bar S^1,\bar S^3\} =  -2i\bar S^0\bar S^2 + {\cal O}(q)
\nn \\
\{\bar S^2,\bar S^3\} = 2i\bar S^0\bar S^1 + {\cal O}(q)
\ee
with the Lax matrix
\be\label{laxds}
L_{ds}=\bar S^0{\bf 1}+i\bar S^3\sigma_3+{i\over
2}\left(\sqrt{w}+{\f\sqrt{w}}\right)\bar S^1\sigma_1 +
{i\over
2}\left(\sqrt{w}-{\f\sqrt{w}}\right)\bar S^2\sigma_2
\ee
One can notice that (\ref{laxxxz}) and
(\ref{laxds}) are essentially the same. In particular,
Lax operator (\ref{laxds})
satisfies the quadratic Poisson relation (\ref{quadrP}) with the same
trigonometric $r$-matrix (\ref{trigrmatrix}). Indeed, these two Lax matrix are
related by the simple transformation
\be
L_{ds}=-\sin\left(\pi\xi\sigma_2\right)L_{XXZ}
\ee
$w$ being identified with $e^{2i\xi}$ and $\bar S^0$, $\bar S^1$, $\bar S^2$,
$\bar S^3$ with $\hat S^2$, $\hat S^3$, $\hat S^0$,
$\hat S^1$ respectively. Let us also note
that Lax operator (\ref{laxds}) is nothing but the $L$-operator of the
lattice Sine-Gordon model.

The determinant
$\det _{2\times 2} \hat L(\xi)$ is equal to
\be
\det _{2\times 2} \hat L(\xi) = \hat S_0^2 + \sum_{a=1}^3 e_a\hat S_a^2
- \wp(\xi)\sum_{a=1}^3\hat S_a^2 = \nn \\
= K - M^2\wp(\xi) =  K - M^2 x
\ee
where
\be\label{casi}
K = \hat S_0^2 + \sum_{a=1}^3 e_a(\tau)\hat S_a^2 \ \ \
\ \ \ \
M^2 =  \sum_{a=1}^3 \hat S_a^2
\ee
are the Casimir operators of the Sklyanin algebra (i.e. Poisson
commuting
with all the generators $\hat S^0$, $\hat S^1$, $\hat S^2$,
$\hat S^3$). The determinant of the monodromy matrix (\ref{monomat})
is
\be\label{54}
Q(\xi) = \det _{2\times 2} T_{N_c}(\xi) =
\prod_{i=1}^{N_c} \det _{2\times 2} \hat L(\xi - \xi_i) =
\prod_{i=1}^{N_c} \left( K_i - M^2_i\wp(\xi - \xi_i)\right)
\ee
while the trace ${\cal P}(\xi) = \frac{1}{2}\Tr T_{N_c}(\xi)$
generates mutually Poisson-commuting Hamiltonians, since
\be
\left\{\Tr T_{N_c}(\xi),\
\Tr T_{N_c}(\xi') \right\} = 0
\ee
For example, in the case of the {\it homogeneous} chain (all $\xi_i = 0$
in (\ref{54}))
$\Tr T_{N_c}(\xi)$ is a combination of the
polynomials:
\be
{\cal P}(\xi) = Pol^{(1)}_{\left[\frac{N_c}{2}\right]}(x) +
y Pol^{(2)}_{\left[\frac{N_c-3}{2}\right]}(x),
\ee
where $\left[\frac{N_c}{2}\right]$ is integral part
of ${N_c\over 2}$, and the coefficients of $Pol^{(1)}$ and $Pol^{(2)}$
are Hamiltonians of the $XYZ$ model
\footnote{For the {\it inhomogeneous} chain the explicit expression for the
trace is more sophisticated: one should make use of the
formulas like
$$
\wp(\xi - \xi_i) = \left(\frac{\wp'(\xi) + \wp'(\xi_i)}
{\wp(\xi) - \wp(\xi_i)}\right)^2 - \wp(\xi) - \wp(\xi_i) =
4\left(\frac{y+y_i}{x-x_i}\right)^2 - x - x_i
$$
}.
As a result, the spectral equation (\ref{fsc-sc2}) for the $XYZ$ model
acquires the form:
\be\label{specXYZ}
\tilde w + \frac{Q(\xi)}{\tilde w} = 2{\cal P}(\xi),
\ee
where for the {\it homogeneous} chain ${\cal P}$ and $Q$
are polynomials in $x = \wp(\xi)$
and $y =   \frac{1}{2}\wp'(\xi)$.
Eq. (\ref{specXYZ}) describes the double covering of the elliptic
curve $E(\tau)$:
with generic point $\xi \in E(\tau)$ one associates the
two points of ${\cal C}^{XYZ}$, labeled by two roots $w_\pm$
of equation (\ref{specXYZ}).
The ramification points correspond to
$\tilde w_+ = \tilde w_- = \pm\sqrt{Q}$, or
$Y = \frac{1}{2}\left(\tilde w - \frac{Q}{\tilde w}\right) =
\sqrt{{\cal P}^2 - Q}= 0$.

The curve (\ref{specXYZ}) is in fact similar to that of $N_c=2$
Calogero-Moser system (\ref{caln2}).
The difference is that now $x = \infty$ is {\it not} a branch point,
therefore, the number of cuts on the both copies of $E(\tau)$ is $N_c$ and the
genus of the spectral curve is $N_c+1$.

Rewriting analytically ${\cal C}^{XYZ}$ as a system of equations
\be\label{cxyz}
y^2 = \prod_{a=1}^3 (x - e_a), \nn \\
Y^2 = {\cal P}^2 - Q
\ee\label{hdb}
the set of holomorphic 1-differentials on ${\cal C}^{XYZ}$ can be chosen as
\be
v = \frac{dx}{y},\ \ \\
V_\alpha = \frac{x^\alpha dx}{yY} \ \ \
\alpha = 0,\ldots,
\left[\frac{N_c}{2}\right], \nn \\
\tilde V_\beta = \frac{x^\beta dx}{Y} \ \ \
\beta = 0, \ldots,
\left[\frac{N_c-3}{2}\right]
\ee
with the total number of holomorphic 1-differentials
$1 + \left(\left[\frac{N_c}{2}\right] + 1\right) +
\left(\left[\frac{N_c-3}{2}\right] + 1\right) = N_c+1$
being equal to the genus of ${\cal C}^{XYZ}$.

\section{Generating 1-form}

Given the spectral curve and the integrable system one can immediately
write down the ``generating" 1-differential
$dS$ which obeys the basic defining property
(\ref{basic}). For the Toda chain it can be chosen in two different ways
\be\label{dstc}
d\Sigma^{TC} \cong d\lambda\log w \ \ \ \ \  dS^{TC} \cong \lambda{dw\over w}
\nn \\
d\Sigma ^{TC} = - dS^{TC} + df^{TC}
\ee
Both $d\Sigma ^{TC}$ and $dS^{TC}$ obey the basic property (\ref{basic}) and,
while $f^{TC}$ itself is {\it not} its variation, $\delta f^{TC} =
\lambda{\delta w\over w}$ appears to be a (meromorphic) single-valued
function on ${\cal C}^{TC}$.

In the $XXX$ case \cite{GMMM}, one has almost the same formulas as
(\ref{dstc})
\be\label{dsxxx}
d\Sigma^{XXX} \cong d\lambda\log w \ \ \ \ \
dS^{XXX} \cong \lambda{dw\over w} \nn \\ d\Sigma ^{XXX} = - dS^{XXX} +
df^{XXX} \nn \\ w = {\tilde w\over\sqrt{\det T_{N_c}(\lambda )}}
\ee
For the
$XYZ$ model (\ref{specXYZ}) the generating 1-form(s) $dS^{XYZ}$ can be
defined as
\be\label{dsxyz}
d\Sigma ^{XYZ} \cong d\xi \cdot \log w
\nn \\
dS^{XYZ} \cong \xi {dw\over w} = - d\Sigma ^{XYZ} + d(\xi\log w)
\ee
Now, under the variation of moduli (which are all contained in ${\cal P}$,
while $Q$ is moduli independent),
\be
\delta(d\Sigma ^{XYZ}) \cong \frac{\delta w}{w}d\xi =
\frac{\delta{\cal P(\xi)}}{\sqrt{{\cal P(\xi)}^2-Q(\xi)}} d\xi
= \frac{dx}{yY}\delta{\cal P}
\ee
and, according to (\ref{hdb}), the r.h.s. is a {\it holomorphic}
1-differential on the spectral curve (\ref{specXYZ}).

The singularities of $d\Sigma ^{XYZ}$ are located at the points where $w=0$ or
$w=\infty$, i.e.  at zeroes of $Q(\xi)$ or poles of ${\cal P}(\xi)$.
In the vicinity of a singular point, $d\Sigma ^{XYZ}$ is not
single-valued but acquires addition $2\pi id\xi$ when circling around this
point.
The difference between $d\Sigma$ and $dS$ is again a total derivative,
but $\delta f^{XYZ} = \xi{\delta w\over w}$ is not a single-valued function.
In contrast to $d\Sigma ^{XYZ}$,
$dS^{XYZ}$ has simple poles at $w=0,\infty$ with the residues
$\left.\xi\right|_{w=0,\infty}$,
which are defined modulo $1,\tau$. Moreover, $dS^{XYZ}$ itself is
multivalued: it changes by $(1,\tau)\times\frac{dw}{w}$ when circling
along non-contractable cycles on $E(\tau)$.

Naively, neither $d\Sigma ^{XYZ}$ nor $dS^{XYZ}$ can play the role of the
Seiberg-Witten
1-form -- which is believed to possess well-defined residues, interpreted as
masses of the matter hypermultiplets \cite{SW2}.

In the simplest example $N_c=2$, the second equation in (\ref{cxyz}) is
\be\label{nc2cur}
Y^2 = {\cal P}^2 - Q = (H_0 - H_2x)^2 - ( K_1 -  M^2_1x)
(K_2- M^2_2x)
\nn \\
\equiv  A(x - x_1)(x - x_2)
\ee
It is a curve of genus $N_c+1 = 3$, obtained by gluing two
copies of $E(\tau)$ along two cuts:  between $x=x_1$ and $x=x_2$ on every
of two sheets of $E(\tau)$.  In (\ref{nc2cur})
\be\label{hams}
H_0 = \hat
S^0_1\hat S^0_2 + \sum_{a=1}^3 e_a\hat S^a_1\hat S^a_2 \ \ \ \ \ \ H_2 =
\sum_{a=1}^3 \hat S^a_1\hat S^a_2
\ee
and, comparing with (\ref{casi}),
it is natural to represent
\be H_2 = M_1 M_2 \cos h
\ee
Such a separation of
the Casimir ($M$) and moduli ($h$) dependence is implied by considering the
various limits: conformal one -- with all $M_i \rightarrow 0$ and a
``dynamical transmutation" regime when some $M_i\rightarrow\infty$ along
with $\tau \rightarrow +i\infty$.

When $\tau \rightarrow + i\infty$ or $q = e^{i\pi\tau}
\rightarrow 0$,
the ramification points $e_1$ and $e_2$ collide: $e_1 - e_2 = 16q +
{\cal O}(q^3)$
and the proper coordinates on ${\cal C}^{XYZ}$ are
$ x = -\frac{1}{3} + q\check x$, $y = q\check y$.
Then, equation (\ref{ell}) for $E(\tau)$ turns into:
\be
\check y^2 = \check x^2 - 1
\ee
describing the double-sheet covering of $CP^1$, which is again
$CP^1$. The canonical holomorphic 1-form $d\xi = 2\frac{dx}{y}$
turns into
\be
2\frac{d\check x}{\check y} = 2\frac{d\check x}{\sqrt{\check x^2-1}} =
2\frac{dz}{z}
\ee
where $\check x = z + z^{-1}$.

The {\it double}-scaling limit assumes that the ramification points
$x_1$ and $x_2$ also behave in a special way as $q\rightarrow 0$.
Namely, let
\be\label{scalingA}
x_i = -\frac{1}{3} + q\check x_i
\ee
Now, rescaling $Y = q\check Y$, one gets for ${\cal C}^{XYZ}$
in the double-scaling limit
\be
\check y^2 = \check x^2 - 1, \nn \\
\check Y^2 =  A(\check x - \check x_1)(\check x - \check x_2)
\ee
These equations describe two copies of $CP^1$ glued together along the
two cuts (between $\check x = \check A_1$ and $\check x = \check A_2$
on every of two sheets) -- i.e. this is an elliptic curve (torus)
of genus 1.

The generating 1-form
\be
d\Sigma ^{XYZ} \cong d\xi \cdot \log w
\rightarrow d\Sigma ^{TC} \cong \frac{dz}{z}\log w
\ee
For generic $N_c$ the multi-scaling limit can be performed in a similar to
(\ref{scalingA}) way implying that the full spectral curve of
genus $N_c+1$ -- a double covering of $E(\tau)$ -- degenerates
into the double-covering of $CP^1$, which is of genus $N_c-1$
and is associated with the Toda-chain system. The generating differentials
$d\Sigma ^{XYZ}$ and $dS^{XYZ}$ also
turn into the corresponding Toda-chain generating 1-forms (\ref{dstc}).

\section{Comments}

We discussed some elementary results about the $XYZ$ chain from the
perspective of the Seiberg-Witten exact solutions. In this framework,
one associates with every (finite-gap) integrable system a family of
spectral curves (so that the integrals of motion play the role of
moduli of the complex structures)
and (the cohomology classes of) the generating 1-form
$dS$, satisfying (\ref{basic}), which can be used to construct
the ``periods'' $a$, $a_D$ and the prepotential.
According to \cite{GKMMM}, the prepotential for the $4d$
${\cal N}=2$ SQCD coincides with (the logarithm of) the $\tau $-function of
the associated integrable model.

One can naturally assume that the $XYZ$ chain, which is an
elliptic deformation of the $XXX$ chain known
to describe ${\cal N}=2$  supersymmetric QCD with $N_f < 2N_c$ \cite{GMMM},
can be associated with the  $N_f = 2N_c$ case.
This would provide a description of the conformal (UV-finite)
supersymmetric QCD, differing from the conventional one \cite{SW2,FuMa}.
However, as demonstrated in the present text,
there are several serious differences between $XYZ$
and $XXX$ models, which should be kept in mind.
Let us list some of them.

{\bf 1)} Normally, there are two natural ways to introduce
$d^{-1}(symplectic\ form)$: as $pdq$ and $qdp$ -- these are
represented by the meromorphic 1-differentials $dS$ and $d\Sigma$
in the main text.
Usually, {\it both} satisfy (\ref{basic}); $d\Sigma$
has no simple poles, but is not single-valued on ${\cal C}$,
while $dS$ is single-valued and possesses simple poles
(of course, in general $dS \ncong d\Sigma$). The proper generating
1-form is $dS$.
However, in the $XYZ$ case, {\it both} $dS$ and $d\Sigma$ are {\it not}
single-valued. Moreover, the residues of $dS$ -- identified
with masses of the matter hypermultiplets in the framework of \cite{SW2}
-- are defined only modulo $(1,\tau)$.

{\bf 2)} As $\tau \rightarrow i\infty$, the $XYZ$ model turns into
$XXZ$ rather than $XXX$ one. This makes the description of the
``dimensional transmutation'' regime rather tricky.

{\bf 3)} Starting from the spectral curve (\ref{SpeC}) for the
Toda-chain (pure ${\cal N}=2$ SYM), the Calogero-Moser deformation is
associated with ``elliptization" of the $w$-variable, while
the $XYZ$-deformation -- with that of the $\lambda$-variable.
It is again nontrivial to reformulate the theory in such
a way that the both deformations become of the same nature.
One of the most naive pictures would associate the Hitchin-type
(Calogero) models with the ``insertion" of an $SL(N_c)$-orbit at
one puncture on the elliptic curve, while the
spin-chain ($XYZ$) models -- with the $SL(2)$-orbits
at $N_c$ punctures. Alternatively, one can say that the
$XYZ$-type ''elliptization", while looking local (i.e. $L$-operators
at every site are deformed independently), is in fact a global
one (all the $L$'s can be elliptized  only simultaneously, with
the same $r$-matrix and $\tau$), --  but this is not clearly
reflected in existing {\it formalism}, discussed in this paper.
Moreover, the proper formalism should naturally allow one
to include any simple Lie group (not only $SL(N_c)$)
and any representations (not only adjoint or fundamental).

Already these comments are enough to demonstrate that the hottest
issues of integrability and quantum-group theory (like notions
of elliptic groups and dynamical $R$-matrices) can be of immediate
importance for the Seiberg-Witten (and generic duality) theory.
These subjects, however, remain beyond the scope of this note.

\section{Acknowledgements}

We are indebted to
O.Aharony, G.Arutyunov, S.Gukov, A.Hanany, H.Itoyama, S.Khar\-chev,
I.Kri\-chever, I.Po\-lyubin, V.Rubtsov, J.Sonnen\-schein, B.Voronov and
A.Zabrodin for usefull discussions.

The work of A.G. is supported in part by the grants RFFI-96-01-01101,
INTAS-93-2494, that of A.Mar. by  RFFI-96-01-01106, INTAS-932058 and
that of A.Mir. by RFFI-96-02-19085, INTAS-93-1038 and Volkswagen Stiftung.


\bigskip

\begin{thebibliography}{12}

\bibitem{GMMM}
A.Gorsky, A.Marshakov, A.Mironov and A.Morozov, hepth/9603140
\bibitem{GKMMM}
A.Gorsky, I.Krichever, A.Marshakov, A.Mironov and A.Morozov,
hepth/9505035, Phys.Lett. {\bf B355} (1995) 466
\bibitem{MWNT}
E.Martinec and N.Warner, hepth/9509161 \\
T.Nakatsu and K.Takasaki, hepth/9509162\\
T.Eguchi  and S.Yang, hep-th/9510183
\bibitem{intcal}
R.Donagi and E.Witten, hepth/9510101 \\
E.Martinec, hepth/9510204 \\
A.Gorsky and A.Marshakov, hepth/9510224 \\
E.Martinec and N.Warner, hepth/9511052 \\
H.Itoyama and A.Morozov, hepth/9511126\\
H.Itoyama and A.Morozov, hepth/9512161\\
H.Itoyama and A.Morozov, hepth/9601168
\bibitem{M}
A.Marshakov, hepth/9602005
\bibitem{AN}
C.Ann and S.Nam, hepth/9603028
\bibitem{NT2}
T.Nakatsu and K.Takasaki, hepth/9603069
\bibitem{SW1}
N.Seiberg and E.Witten, Nucl.Phys. {\bf B426} (1994) 19
\bibitem{SW2}
N.Seiberg and E.Witten, hepth/9408099, Nucl.Phys. {\bf B431} (1994) 484
\bibitem{FuMa}
A.Hanany and Y.Oz, hepth/9505075\\
P.Argyres, D.Plesser and A.Shapere, hepth/9505100\\
P.Argyres and A.Shapere, hepth/9509175\\
A.Hanany, hepth/9509176\\
J.Minahan and D.Nemeschansky, hepth/9507032
\bibitem{Skl} E.Sklyanin, J.Sov.Math. {\bf 47} (1989) 2473\\
E.Sklyanin, solv-int/9504001
\bibitem{FT} L.Faddeev and L.Takhtadjan, {\sl Hamiltonian Approach to the
Theory of Solitons}, 1986
\bibitem{KriZa}
I.Krichever and A.Zabrodin,  hepth/9505039
\bibitem{JapBob}
E.Date, M.Jimbo, M.Kashiwara and T.Miwa, Publ. RIMS-395 (1982) \\
A.Bobenko, Func.Anal. \& Apps. {\bf 19} (1985) 6\\
E.Belokolos, A.Bobenko, V.Matveev and V.Enolsky, Usp. Mat. Nauk. {\bf 41}
(1986) 3
\bibitem{Hit}
B.Dubrovin, I.Krichever and S.Novikov, {\it Integrable systems - I},
{\sl Sovremennye problemy matematiki (VINITI), Dynamical systems - 4}
(1985) 179
N.Hitchin, Duke.Math.Jour. {\bf 54} (1987) 91 \\
E.Markman, Comp.Math. {\bf 93} (1994) 255\\
A.Gorsky and N.Nekrasov, hepth/9401021
\bibitem{NekRub}
N.Nekrasov, hepth/9503157\\
B.Enriquez, V.Rubtsov, alg-geom/9503010
\bibitem{Jimbo} M.Jimbo, Comm.Math.Phys. {\bf 102} (1986) 537
\bibitem{KriW}
I.Krichever, Func.Anal.\& Appl. {\bf 22} (1988), 37
\bibitem{KriDu}
I.Krichever, Uspekhi Mat.Nauk {\bf 33:4} (1976) 215\\
I.Krichever,  in the Appendix to B.Dubrovin,  Uspekhi Mat.Nauk
{\bf 36:2} (1981) 12
\bibitem{KriCal}
I.Krichever, Func.Anal. \& Appl.{\bf 14} (1980) 282
\bibitem{RCal}
E.Sklyanin, hepth/9308060, Alg.Anal. {\bf 6} (1994) 227 \\
H.W.Braden and T.Suzuki, hepth/9309033, Lett.Math.Phys. {\bf 30} (1994) 147\\
G.Arutyunov and P.Medvedev, hepth/9511070
\bibitem{Ino}
V.Inozemtsev, Comm.Math.Phys. {\bf 121} (1989) 629
\bibitem{Skl1}
E.Sklyanin, Func.Anal \& Apps. {\bf 16} (1982) 27\\
E.Sklyanin, Func.Anal \& Apps. {\bf 17} (1983) 34


\end{thebibliography}
\end{document}